\documentclass[a4paper,11pt] {article}
\usepackage{times,graphicx}
\usepackage{xspace}
\usepackage[paper=a4paper,dvips,top=2.5cm,left=3cm,right=3cm, foot=1cm,bottom=2cm]{geometry}

\newcommand{\bua}{Bottom-Up Approach\xspace}
\newcommand{\MET}{${\rm E}_{\rm T}^{\rm miss}$}

\begin{document}

\title{A \bua to SUSY Analyses}

\author{Claus Horn \\
\it SLAC National Accelerator Laboratory, Stanford University, CA\\
}
\date{}

\maketitle

\begin{abstract}
This paper proposes a new way to do event generation and analysis in searches for new physics at the LHC. 
An abstract notation is used to describe the new particles on a level which better corresponds to detector resolution of LHC experiments. In this way the SUSY discovery space can be decomposed into a small number of eigenmodes each with only a few parameters,
which allows to investigate the SUSY parameter space in a model-independent way. By focusing on the experimental observables for each process investigated
the Bottom-Up Approach allows to systematically study the boarders of the experimental efficiencies and thus to extend the sensitivity for new physics.
\end{abstract}

\begin{section}{Introduction}
To be prepared for the potential discovery of all   
phenomenological manifestations of the 120 dimensional parameter space of the MSSM
poses challenges to computing and manpower even in modern day collaborations.
Experimental approaches in the past mostly focused on a small number of 
theoretically motivated benchmark points or generated grids of points in simplified 
models with less parameters but also reduced phenomenological coverage.
Newer ideas generate large numbers (order of 1000) of random points in higher dimensional spaces.

Former approaches to general searches for new physics tried to look for all possible combinations of final state particles \cite{Knuteson}
and did not take into account that these signatures are the result of an underlying structure induced by the fundamental interactions of the particle types of the MSSM 
which can be classified as presented here.
These structures (the elementary mass spectra of Sect. \ref{sec:elemetarymassspectra})   
are the main source of the observed correlations and degeneracies in the mapping between SUSY parameters and observable signatures.
Therefore, solving the inverse problem \cite{inverseProblem} does not start with inclusive measurements of any given final state
but it starts with doing the analyses according to these structures.

The approach presented here will argue that the only experimentally relevant parameters are the masses of the new particles,
resulting in a very small number of parameters that have to be considered in the analysis of each eigenmode (the dominant observable channels).
The aim of each analysis in this approach is to map out the detector efficiency as a function of these parameters.
The mapping from observables to the parameters of a theoretical model
can then be factored out to the generator level, omitting the time consuming generation and analysis of events for
each parameter point. 
As a result, the experimental findings can be interpreted in many different theoretical models in many dimensions and with high precision. 

The idea to focus directly on the parameters of the new particles instead of the abstract parameters of a particular SUSY model 
has been applied to specific problems in the past (see Sect. \ref{sec:examples})
and has inspired the proposal for an investigation of simplified SUSY models \cite{simplifiedModels}.
Unfortunately, it has not been applied in LHC analyses so far.

The point made in this paper is that searches for new physics in general could benefit significantly from a Bottom-Up analysis approach. 
In particular, it is argued that an application of the \bua would allow to cover the SUSY discovery space in a model independent way.
\end{section}

\begin{section}{Observables}
New SUSY particles are characterized by a number of quantum numbers, of which most, like e.g. the spin, are 
fixed for a given particle type and do not vary with the SUSY model parameters.
The only quantities that may vary are the masses and effective couplings of the new particles.

While special SUSY breaking models pose restrictions on the mass ratios that can be investigated
and therefore cover only parts of the observable space.
The \bua allows to investigate the complete observable space accessible to experiment at once and in a systematic way
by directly varying the masses of the new particles in the events generated.

On the other hand, the couplings only change the branching ratios and thus have no effect on the efficiency of each elementary mass spectrum analysis. 
Similarly, all theoretical distinctions in the particles, for instance between $u_R$ and $u_L$, and theoretical parameters 
which only affect the branching ratios, like 
the trilinear couplings $A_i$, are irrelevant for the event generation and analysis step 
and can be factored out to a separate step of theoretical interpretation.
The experimentally relevant parameters that have to be considered in event generation are
\footnote{This may be extended by considering a third neutralino $\chi_3$. However, the findings in Sect. \ref{sec:pmssmana} indicate that its contribution is probably very small.}:
\begin{equation}
    m(\tilde{g}), \{ m(\tilde{q}), m(\tilde{b}), m(\tilde{t}) \}, m(\tilde{\chi_1}), m(\tilde{\chi_2}), \{ m(\tilde{e}), m(\tilde{\mu}), m(\tilde{\tau}), m(\tilde{\nu}) \}. 
\end{equation}
However, by decomposing the analysis according to their eigenmodes, only up to five of them are relevant at once for the analysis of each mode (as indicated by the brackets).

In addition, the experiments will deliver measurements of the effective cross sections for each channel 
investigated. These can be compared to determine the branching fractions and thus indirectly provide measurements 
for the couplings of the new particles and finally the composition of the gauginos which might 
shed some light on the nature of the underlying SUSY model.
\end{section}

\begin{section}{Elementary Mass Spectra of the MSSM} \label{sec:elemetarymassspectra}
Each complex mass spectrum (comprising all sparticles found at a given SUSY model point) can be decomposed into a small number of elementary mass spectra by applying the following transformations:
\begin{itemize}
\item Convert it to Abstract Notation \cite{myThesis} by replacing all squarks with $\tilde{q}$, all gauginos with $\tilde{\chi}$ and all sleptons with $\tilde{l}$;\footnote{Note that this level of abstraction preserves the two main features of the new particles: Their spins (which determine angular distributions) and their types of interactions (which determine the types of standard model particles they may decay into).}
\item Identify the decay channels with the highest value in ${\rm BR} \frac{\epsilon_S}{\epsilon_B}$.\footnote{BR denotes the branching ratio and $\epsilon_S$, $\epsilon_B$ the selection efficiencies for signal and background events, respectively. 
}  
\end{itemize} 
As a result one recovers always the same small number of elementary mass spectra for all MSSM parameter points\footnote{One complex mass spectrum may comprise several versions of the same elementary mass spectrum.}.

To see which are the elementary mass spectra of the MSSM one may attempt to 
construct them from scratch.
In the construction the following constraints have to be fulfilled:
\begin{itemize}
\item To result in a high cross section, sparticles have to be produced via strong couplings, i.e. via $\tilde{q}\tilde{q}$, $\tilde{q}\tilde{g}$ or $\tilde{g}\tilde{g}$ production;
\item To ensure a cold dark matter candidate the LSP can be only $\tilde{\chi}_1^0$ or $\tilde{G}$;
\item Due to phase space constraints, the branching ratios for longer decay chains are very small. The only exceptions that have to be considered are:
  \begin{itemize}
  \item Wino like higher gauginos, or
  \item Intermediate sleptons which increase the selection efficiency. 
\end{itemize}
\end{itemize}
\begin{figure*}[t]
  \centering
\includegraphics[width=12cm]{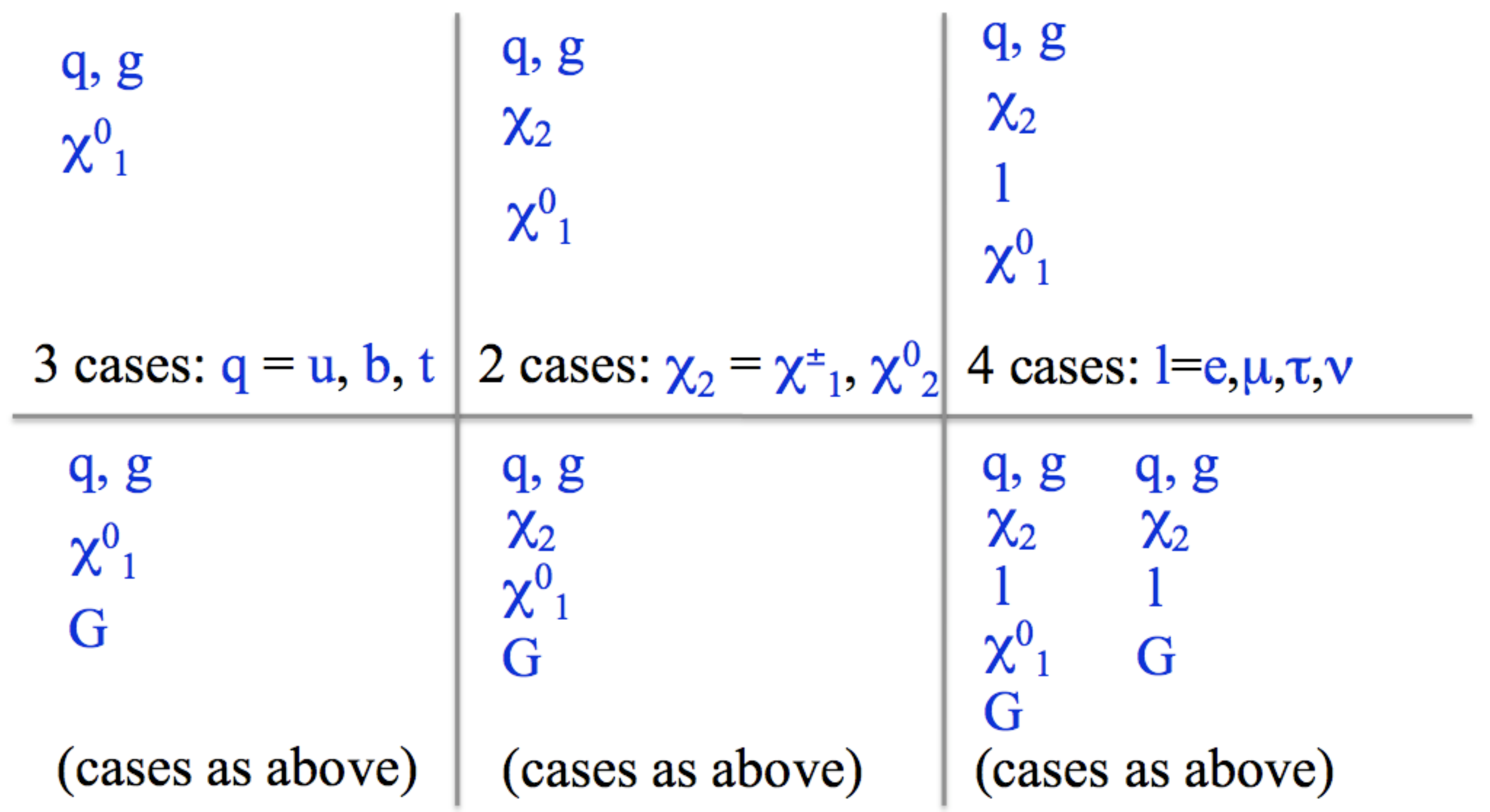}
  \caption{Elementary mass spectra of the MSSM. Shown are sparticles in abstract notation. G denotes the gravitino.}
  \label{fig:elementarymassspectraMSSM}
\end{figure*}
A set of elementary mass spectra that can be derived in this way is shown in Figure \ref{fig:elementarymassspectraMSSM}.
Together they form the MSSM discovery space.
Since experiments allow to differentiate between light, heavy-flavor and top quarks as well as different types of leptons they may be subdivided into different cases as indicated in the figure
\footnote{Note that some longer spectra, like the three gaugino case, could eventually be added but, as will be seen in Sect. \ref{sec:pmssmana}, they can be expected to have minor effects on the discovery potential.
Also note that additional subdivisions are possible but they are considered internal to an analysis
and do not affect the outcome of the discussion.}.

Note that Fig. \ref{fig:elementarymassspectraMSSM} provides a compact overview of possible SUSY signatures,
not all of which are covered by current LHC analyses. One observes, for instance, that isolated, high energy photons are very important and 
appear in combination with many other final states. 
\end{section}

\begin{section}{Analysis of pMSSM Points} \label{sec:pmssmana}
As a cross check for the sensibility of the results obtained in the last section  
1000 randomly generated and not yet excluded pMSSM \cite{pMSSMparameters} 
parameter points (see \cite{JoAnneTom} for a description of the experimental constraints applied) were investigated with respect to their elementary mass spectra content.
The steps performed were:
\begin{itemize}
\item PYTHIA was used to derive the complex mass spectra and decay tables for all sparticles for each pMSSM parameter point\footnote{The use of PYTHIA might not always give the exact result and it was not used in \cite{JoAnneTom}. 
The point of the discussion here is just a qualitative one.
For more detailed and more exact results about the pMSSM see \cite{JoAnneTom}.};
\item Starting with squarks and gluino, all the decay products were recorded and were followed iteratively until the LSP, resulting typically in ten to forty possible decay chains for each pMSSM point;
\item Each decay chain was translated into abstract notation and the corresponding elementary mass spectra were extracted, summing the contributing branching ratios, resulting typically in about one to four elementary mass spectra for each pMSSM point;
\item The average and maximum branching ratios were calculated for all 1000 pMSSM points.
\end{itemize}

The result is shown in Table \ref{tab:elementaryMassSpectra}. Note that the pMSSM points only include spectra where the $\chi_1^0$ is the LSP.
The elementary mass spectra were not distinguished with respect to the number of jets.  
Only decays with ${\rm BR}\ge{1\%}$ were considered for each sparticle.
\begin{table*}
  \centering
  \begin{tabular}{| l | r | r |}
  Elementary Mass Spectra & avg BR [\%] & max BR [\%] \\
  \hline
    $(\tilde{q},\tilde{g}), \tilde{\chi}$ & 86.4477 & 100 \\
    $(\tilde{q},\tilde{g}), \tilde{\chi}, \tilde{\chi}$ & 9.8834 & 100 \\
    $(\tilde{q},\tilde{g}), \tilde{\chi}, \tilde{\nu}, \tilde{\chi}$ & 1.2841 & 67.1513  \\
    $(\tilde{q},\tilde{g}), \tilde{\chi}, \tilde{\tau}, \tilde{\chi}$ & 1.1207 & 42.1953  \\
    $(\tilde{q},\tilde{g}), \tilde{\chi}, \tilde{e}, \tilde{\chi}$ & 0.6217 & 29.5350  \\
    $(\tilde{q},\tilde{g}), \tilde{\chi}, \tilde{\mu}, \tilde{\chi}$ & 0.5578 & 25.7138  \\
    $(\tilde{q},\tilde{g}), \tilde{\chi}, \tilde{\chi}, \tilde{\chi}$ & 0.0840 & 14.4386  \\
    $(\tilde{q},\tilde{g}), \tilde{\chi}, \tilde{\chi}, \tilde{\nu}, \tilde{\chi}$ & 0.0006 & 0.2469 \\
  \end{tabular}
  \caption{Elementary mass spectra of 1000 random not-yet-excluded pMSSM points. ($\tilde{q}$,$\tilde{g}$) denotes any of the productions $\tilde{q}\tilde{q}$, $\tilde{q}\tilde{g}$ or $\tilde{g}\tilde{g}$.}
  \label{tab:elementaryMassSpectra}
\end{table*}

One observes that the number of different elementary mass spectra 
in the pMSSM is indeed quite small and 
in agreement with the arguments given in Section \ref{sec:elemetarymassspectra}.\footnote{Note that contrary to common rumor SUSY decay chains prefer to be short.} 
In contrast to mSUGRA based expectations, the importance of the slepton mode is quite small,
while jet-only and boson modes  
seem to deserve some extra attention.
\end{section}

\begin{section}{Event Generation}
While SUSY analyses traditionally start by choosing a parameter point in a specific 
SUSY model, the \bua starts from the elementary mass spectra derived in Section \ref{sec:elemetarymassspectra}.
This makes it independent of theoretical assumptions of special SUSY models (like the physics at the GUT scale, in a hidden-sector, etc.), 
which are out of the reach of LHC experiments. 
Since the new particles can be characterized on an abstract level just by their 
fundamental interactions, the results may be interpreted on much more general grounds (beyond the existence of supersymmetry).

While traditionally, an analysis is performed on the basis of a complex mass spectrum at some SUSY model point,
the aim in the \bua is the measurement of the effective cross section for a given eigenmode.
Each eigenmode is analyzed separately. 
The combinations of the results from analyzing the different elementary mass spectra then
allows to reconstruct the complex mass spectrum realized by nature.

Hence, for the generation of events the mass spectrum is 
directly specified in the generator (like e.g. PYTHIA)
and only decays to the next lighter sparticle in the spectrum are allowed (For others the branching ratio is set to zero.), 
so that the total branching ratio for the longest decay channel, passing through all sparticles in the spectrum is $100\%$. 

From each PYTHIA input file corresponding to a given eigenmode 
events can then be generated with several different kinematics
by varying the specified mass values  
(see Section \ref{sec:analysis} for more details.).

The decomposition of the analyses according to the elementary mass spectra can be expected to 
be very helpful for the determination of the total sparticle mass spectrum since it separates out the different eigenmodes while the 
traditional approach only looks at the convoluted spectrum at different SUSY model points\footnote{Usually, much information about the masses of the new particles can already be inferred by computing invariant masses and kinematic edges. A dedicated analysis taking into account all kinematic and topological variables and their correlations, however, will lead to improved mass measurements (see e.g. \cite{improvingMassWithEventInfo}).}. 
\end{section}

\begin{section}{Analysis} \label{sec:analysis}
The aim of any search for new particles is to claim discovery, which can be done if the number of expected signal events, $N_{\rm exp}$, exceeds some limit which depends on the number of expected background (B) and observed data events ($N_{\rm obs}$):
\begin{equation}
   N_{\rm exp} > N_{\rm lim}(B, N_{\rm obs})  
\end{equation}
Traditionally, $N_{\rm exp}$ is estimated by generating events for a specific SUSY model point, $\vec{\theta}$, and determining the efficiency at that point, 
$\epsilon(\vec{\theta})$:
\begin{equation}
    N_{\rm exp}(\vec{\theta}) = \mathcal{L} \sum_{\rm ch} \sigma_{\rm ch}(\vec{\theta}) \times \epsilon_{\rm ch}(\vec{\theta}) 
\end{equation}
where $\sigma_{\rm ch}$ is the effective production cross section for a given analysis channel, ch, and $\mathcal{L}$ the integrated luminosity of the data investigated. 
In order to get a good sampling of $N_{\rm exp}$ this requires the analysis of a large number of channels (one for every combination of final states) in a large dimensional space (e.g. ${\rm dim}(\vec{\theta})=19$ in the pMSSM).

Instead, the decomposition proposed here is:
\begin{equation}
 	N_{\rm exp}(\vec{\theta}) = \mathcal{L}  \sum_{\rm em} {\sigma}_{\rm em}(\vec{\theta}) \times \epsilon_{\rm em}[\vec{m}_{\rm em}(\vec{\theta})]
\end{equation}
with a small number of eigenmodes, em (see Sect. \ref{sec:eigenmodes}), and ${\rm dim}(\vec{m}_{\rm em})$ = 2 to 5.

Since it starts from the masses the \bua separates the event generation and analysis step from the interpretation of the results.
In consequence, it allows to profit from the fact that the dominant factor in the calculation of $N_{\rm exp}$, $\sigma_{\rm em}$, which 
varies exponentially with the masses, can be calculated quickly and in high resolution at the generator level.
Therefore, $N_{\rm exp}$ can be determined easily in many dimensions of model parameters while only a few mass spectra variations have to be generated to determine the effect on the detector efficiency, $\epsilon_{\rm em}(\vec{m}_{\rm em})$, which usually varies only by a few percent. 

Thus, the aim of each analysis is to map out the detector sensitivity by generating specific  
efficiency benchmark points in the observable space which are driven by experimental constraints.
This allows the analysis to explicitly focus on the kinematically challenging regions of its specific efficiency space.
Typical examples for such experimental extremes are signatures with very low \MET and jets only (if the LSP is at its lower boarder), 
decay products with very low $p_T$ (if the mass difference between two sparticles is small),
boosted decay products (if the mass difference is large)
and quasi stable particles (if a sparticle's only decay mode offers very little phase space, i.e. the two sparticles have similar masses). 
These borders of efficiency were often neglected in the past resulting in holes in the observable parameter spaces (see e.g. \cite{JoAnneTom}).

For the investigation of the mass parameter space it is important to notice that
the number of necessary points is relatively small 
since the hierarchy has to be conserved. 
For instance, the kinematically extreme low/high mass cases of mass spectra with three new particles can be 
studied by considering just $4$ scenarios (instead of $2^3=8$), as illustrated in Fig \ref{fig:massspectraexample}.\footnote{Note that also mass configurations where some of the intermediate sparticles are off-shell should be considered since they lead to the same final states and have different kinematics. For these cases the number of relevant mass parameters is reduced (see the example in Sect. \ref{sec:susyjets}). }  
\begin{figure*}
  \centering
\includegraphics[width=6cm]{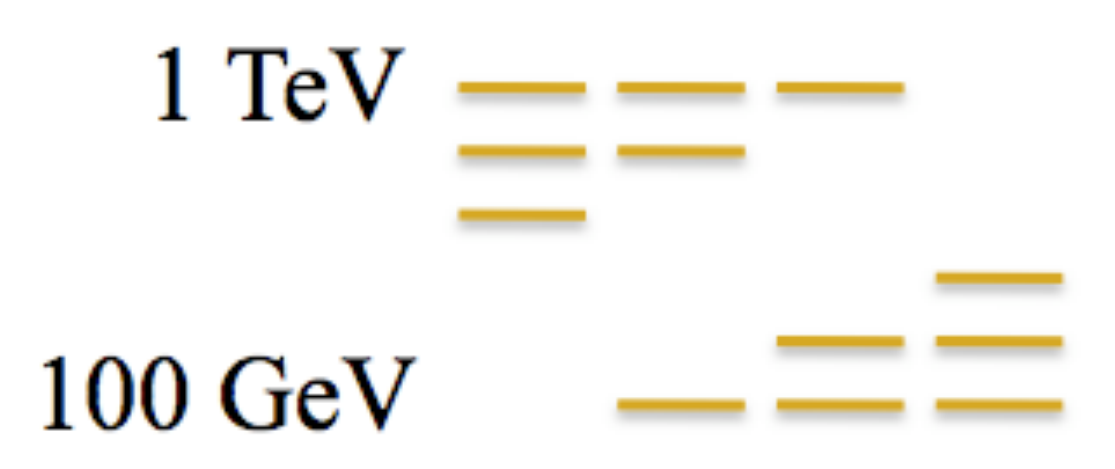}  
  \caption{Example for kinematically extreme cases of mass spectra with three sparticles.}
  \label{fig:massspectraexample}
\end{figure*}
At the level of each analysis it will typically be possible 
to choose a more effective parameterization of the mass space (e.g. by considering mass differences) which allows to concentrate on 
the most important parameters of the detector efficiency.

A variation of the SUSY model parameters will typically result in:
\begin{itemize}
\item No changes of the masses; these parameters are thus irrelevant for analysis, or 
\item Changes of several masses at the same time, thus changing the detector response in a complex way. 
\end{itemize}
A direct variation of the masses, on the other hand, will produce a well understood change in detector response (like the variation of a jet's $p_T$)
and interpolation in the mass space is thus in general straight forward. On the other hand, the change in sensitivity due to changes in model parameters 
is dominated by the effect of changing branching ratios of the different eigenmodes.

Finally, a study of the jet kinematics resulting from variations of the produced squark and gluino masses 
can be factored out since it is common to all elementary mass spectra (see Sect. \ref{sec:susyjets}). 
\end{section}

\begin{section}{Example Applications}\label{sec:examples}
This section describes three brief examples how the \bua may be applied to specific analyses.

\begin{subsection}{Search for Gravitinos}
A search for gravitinos resulting form the GMSB decay $\tilde{\chi} \rightarrow \gamma \tilde{G}$ 
was performed with HERA data where the neutralinos may be produced by the exchange of a virtual slepton \cite{myThesis}.  

A traditional approach of considering different GMSB parameter points
would not allow for the investigation of the complete observable space (due to mass correlations imposed by the GMSB model)
and only allow the analysis of a few GMSB parameter points.
Instead, events were generated for different sparticle masses in agreement with the \bua.
Since the selection efficiency was measured as a function of the neutralino mass (see Figure \ref{fig:myEfficiency}) the results 
could be interpreted in the complete GMSB parameter space by applying the following steps:
\begin{itemize}
\item For each parameter point, the effective signal cross section and neutralino mass was calculated and the kinematic region determined; 
\item The neutralino mass was used to determine the signal efficiency;
\item The efficiency, the effective cross section and the investigated data luminosity were used to calculate the number of expected signal events (similarly for the number of expected background events);
\item  The number of detected data events in the given kinematic region was compared to the number of expected signal and background events to calculate the confidence level for the given parameter point.
\end{itemize}
Note that different selection criteria may be used for different kinematic regions\footnote{In this analysis a multivariate discrimination technique was used which allows for an automatic optimization of the selection criteria for changing signal distributions, see \cite{myThesis} for more details.}. Thus, once the detector efficiency is measured in terms of the relevant parameters all the calculations above only take a few seconds for each parameter point.
\begin{figure*}
  \centering
\includegraphics[width=8cm]{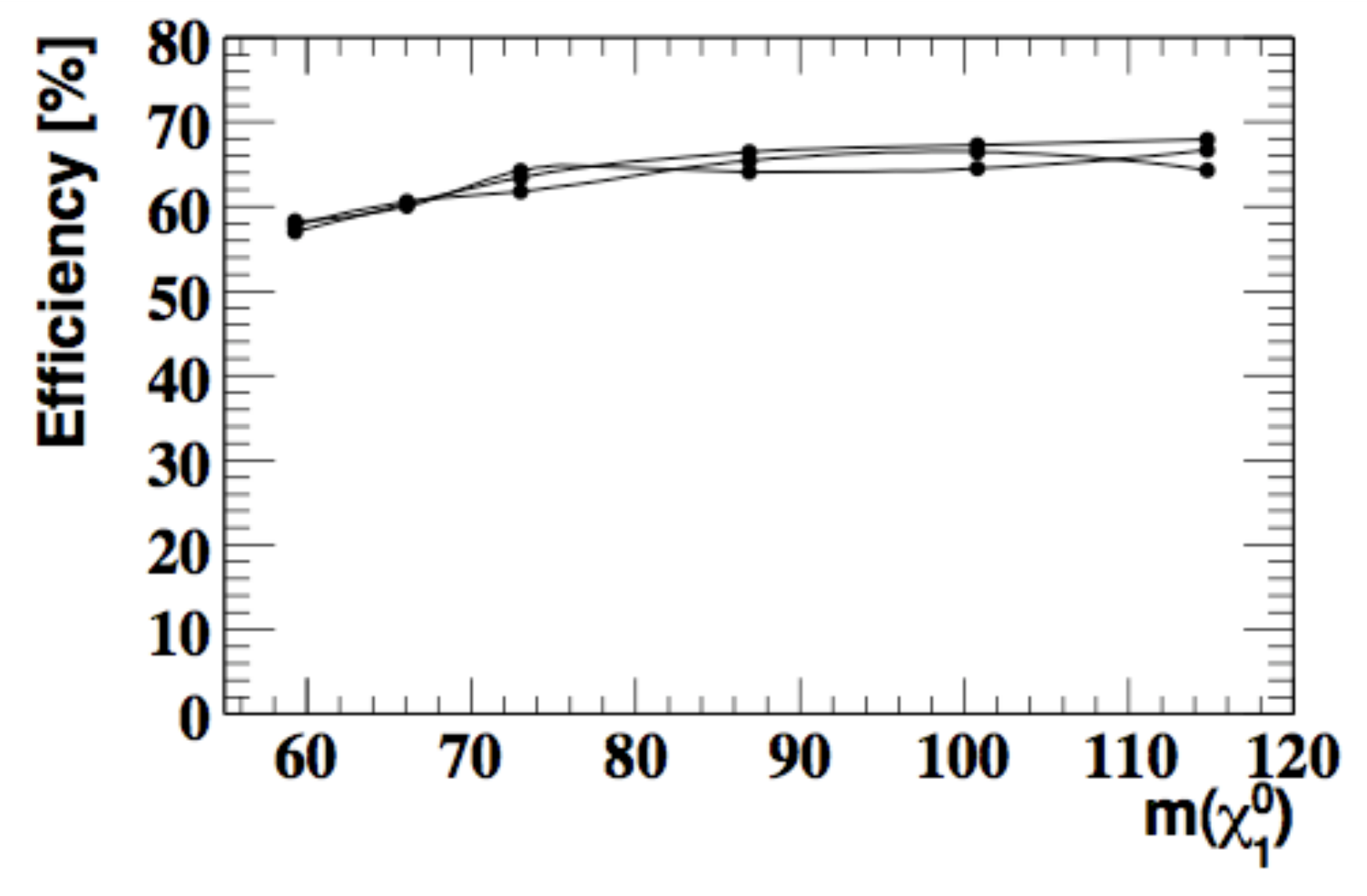}  
  \caption{Selection efficiency as a function of the sparticle masses of the $eq \rightarrow \tilde{e}^* \rightarrow q\tilde{\chi}^0_1 \rightarrow q\gamma\tilde{G}$ process. For each neutralino mass events  were generated for three different slepton masses leading to variations within expected statistical fluctuations. Figure taken from \cite{myThesis}.}
  \label{fig:myEfficiency}
\end{figure*}

{\bf Summary:} While traditionally, the generation and analysis of events for all values of GMSB parameters seems impossible.   
Following the \bua, the complete GMSB parameter space could be investigated by analyzing 
6 different neutralino mass points.
\end{subsection}

\begin{subsection}{SUSY Jets}\label{sec:susyjets}
Traditionally, SUSY discovery reach is often described in the two dimensional mSUGRA plan of $m_0$ and $\,m_{\frac{1}{2}}$. 
This misses the fact that a general description of jet kinematics in SUSY production at the LHC requires three parameters: the masses of the gluino, the lightest squark and the next lightest gaugino. 
However, for different mass combinations a different of the three strong sparticle pair-production processes dominates: 
\begin{itemize}
\item If $m(\tilde{q}) \gg m(\tilde{g})$: $\tilde{g}\tilde{g}$-production dominates leading to events with four parton level jets. 
Jet kinematics depend only on $m(\tilde{g})-m(\tilde{\chi})$.
\item If $m(\tilde{q}) = m(\tilde{g})$: $\tilde{g}\tilde{q}$-production dominates leading to events with three parton level jets.  
\item If $m(\tilde{q}) \ll m(\tilde{g})$: $\tilde{q}\tilde{q}$-production dominates leading to events with two parton level jets. 
Jet kinematics depend only on $m(\tilde{q})-m(\tilde{\chi})$.
\end{itemize}

The first case has been investigated in detail and compared to results from ${\rm D\O}$ \cite{jayGluinoStudy}.
Since the ${\rm D\O}$ analysis was performed in the mSUGRA plane, the reach for a generic $m(\tilde{g})$-$m(\tilde{\chi}_1^0)$ point is unknown (in mSUGRA the gaugino mass ratio is fixed to: $m(\tilde{g})/m(\tilde{\chi}_1^0) = 6$).
Unfortunately, this restricted view hides the fact that different selection criteria would be optimal for different kinematic configurations.
The improved selection criteria proposed in \cite{jayGluinoStudy} are:
\begin{itemize}
\item Optimized cuts for different bins in \MET and ${\rm H}_{\rm T}$ ($=\sum_{\rm jets} E_T$), and
\item In the case where $\tilde{g}$ and $\tilde{\chi}_1^0$ are nearly degenerate, considering cases with initial and final state radiation will result in signatures with increased \MET and thus sensitivity.
\end{itemize}
The result of this study is shown in Figure \ref{fig:gluinoStudy}. 
\begin{figure*}
  \centering
\includegraphics[width=8cm]{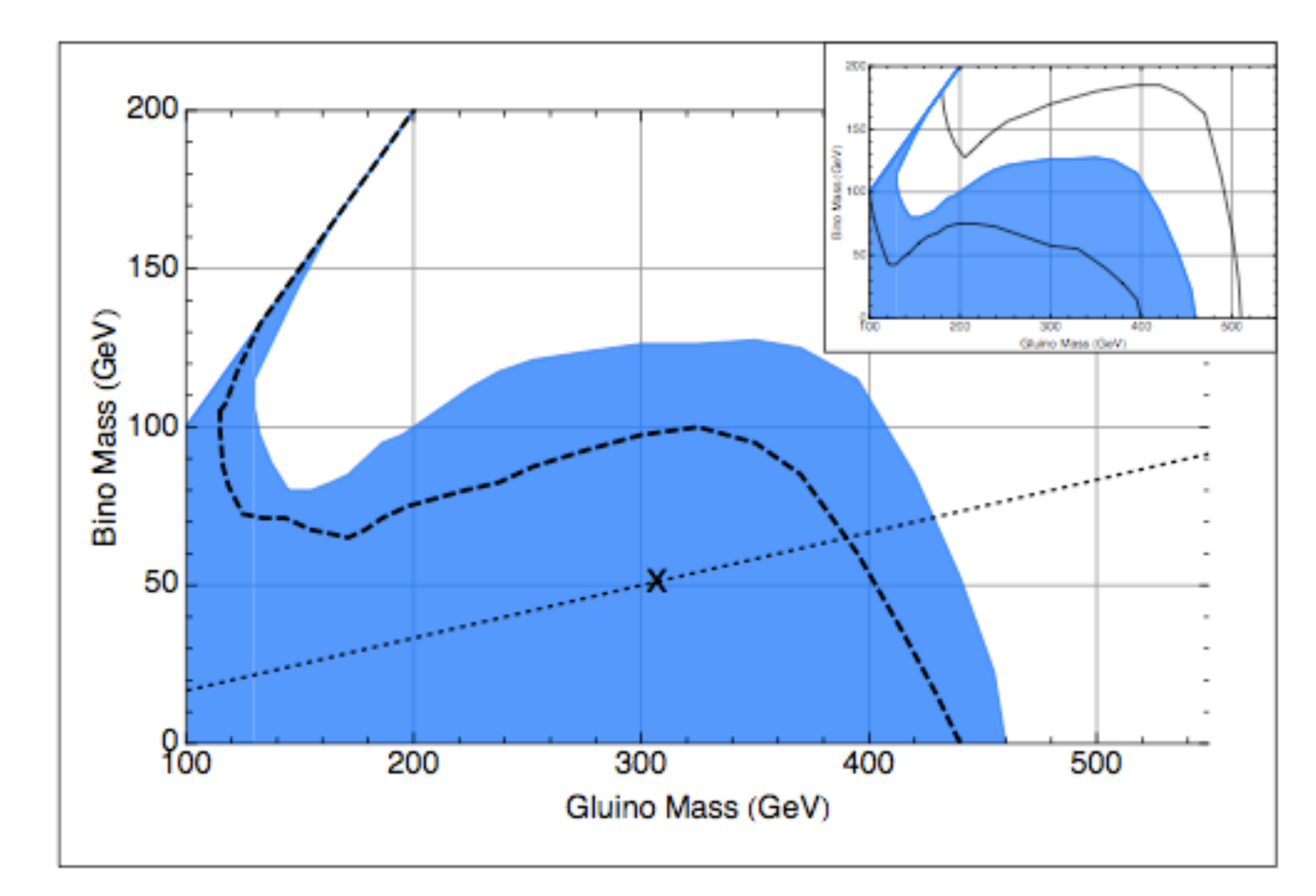}  
  \caption{$m(\tilde{\chi}_1^0)$ versus $m(\tilde{g})$ mass plane. The different symbols depict: Straight dotted line: mSUGRA phase-space; Cross: Current ${\rm D\O}$ limit (for 2.1 ${\rm fb}^{-1}$); dashed-line: expected limit for 4 ${\rm fb}^{-1}$; blue contour: expected limit for improved selection criteria. Figure taken from \cite{jayGluinoStudy}.}
  \label{fig:gluinoStudy}
\end{figure*}
It was shown that by optimizing the selection criteria for different 
mass regions the search reach can be significantly extended.


By adding one more parameter ($m(\tilde{q})$) to the previous example it would be possible to investigate SUSY production at the LHC in a model-independent way.
Thereto, the other two cases ($\tilde{q}\tilde{q}$ and $\tilde{g}\tilde{q}$) should be analyzed in a similar fashion as done for the $\tilde{g}\tilde{g}$ case. 
This would allow to determine the sensitivity for any combination of the three masses (i.e. for any SUSY parameter point)\footnote{The total efficiency is the weighted sum of the efficiencies of the three cases, where the weights are the corresponding production cross sections: $\epsilon_{\rm jets_A} = \frac{1}{\sigma^{\rm gg} + \sigma^{\rm qq} + \sigma^{\rm gq}} (\sigma^{\rm gg}\epsilon_{\rm gg} + \sigma^{\rm qq}\epsilon_{\rm qq} + \sigma^{\rm gq}\epsilon_{\rm gq})$.}. 

More complex analysis with additional sparticles in the spectrum, can then focus on the borders of the so mapped out efficiency space 
to see where it can be extended due to the additionally radiated particles.

{\bf Summary:} The analysis of jet kinematics in the mSUGRA plane leaves holes in the observable parameter space
and does not consider all kinematic configurations.
Following the \bua, a complete study of the SUSY jet kinematics is possible independent of the underlying SUSY model.  
Additionally, it would be possible to define a few jet-benchmark points at the boarder of the trigger efficiency
which could serve as a reference for more complex analyses (see Sect. \ref{sec:eigenmodes}).
\end{subsection}

\begin{subsection}{Boosted Neutralinos}
The ATLAS collaboration has performed a study to identify boosted LSPs decaying into three jets via RPV $\lambda^{"}$ couplings by 
analyzing the jet substructure \cite{boostedNeutralino}.
In order to investigate the sensitivity within some SUSY model space
one would traditionally analyze a number of benchmark points. 

However, following the \bua the efficiency can be determined in a model-independent way by applying the following steps: 
\begin{itemize}
\item Consider the simplest elementary mass spectrum only consisting of a squark and a neutralino.
\item Generate events for a few different mass points to map out the efficiency as a function of neutralino boost and mass.
\item Write a function which returns the distribution of neutralino boosts (using a generator) as a function of the masses of the sparticles in a given decay chain.
\item Chose theoretical parameters to interpret the results in, run a generator to calculate the mass spectrum for each parameter point, determine the neutalino boost and calculate the sensitivity for that parameter point\footnote{The total efficiency is a weighted sum over all decay chains, where the weights are the branching ratios. }.
\end{itemize}
A conservative estimate for the trigger efficiencies can be taken from the studies proposed in the last example. 

{\bf Summary:} While traditionally, the generation and analysis of events for all pMSSM parameters would be impossible.
Following the \bua, the effect of changing kinematics can be determined in a model-independent way and
the sensitivity for any pMSSM parameter point can be calculated at generator level. 
\end{subsection}

\begin{paragraph}
\noindent The above are just a few examples of possible benefits of the \bua.
One can easily imagine how it could profitably be applied to many other analyses 
and especially the eigenmode analyses described in the next section.
\end{paragraph}
\end{section}

\begin{section}{Eigenmodes}\label{sec:eigenmodes}
One additional complication not addressed so far is that
the elementary mass spectra may appear in different combinations  
since sparticles are pair-produced resulting in two separate decay chains per event\footnote{Although this section focuses on the neutralino LSP case, the extension to gravitino LSPs should be straight forward.}.

With the aim to investigate the sensitivity for a given elementary mass spectrum, 
the question arises which of the combinations with the other spectra is prevailing.
For the pMSSM points investigated in Sect. \ref{sec:pmssmana} the answer is shown in Table \ref{tab:prevailance}.
\begin{table*}
  \centering
  \begin{tabular}{| l | c | c | c | }
             & $(\tilde{q},\tilde{g}), \tilde{\chi}$ & $(\tilde{q},\tilde{g}), \tilde{\chi}, \tilde{\chi}$  &  $(\tilde{q},\tilde{g}), \tilde{\chi}, \tilde{l}, \tilde{\chi}$ \\
  \hline
      $(\tilde{q},\tilde{g}), \tilde{\chi}$ & 75\% & 17\% & 6\% \\
      $(\tilde{q},\tilde{g}), \tilde{\chi}, \tilde{\chi}$ & 17\% & 0.9\% & 0.7\% \\
      $(\tilde{q},\tilde{g}), \tilde{\chi}, \tilde{l}, \tilde{\chi}$ & 6\% & 0.7\% & 0.1\% \\
  \end{tabular}
  \caption{Relative importance of elementary mass spectra combinations for the 1000 pMSSM points of Section \ref{sec:pmssmana}.
  }
  \label{tab:prevailance}
\end{table*}
As one might expect, the contribution of each spectrum is biggest when combined with the shortest mode, $(\tilde{q},\tilde{g}), \tilde{\chi}$.
It seems thus reasonable to focus on the three classes of eigenmodes illustrated in Figure \ref{fig:theeigenmodes}.
\begin{figure*}
  \centering
\includegraphics[width=9cm]{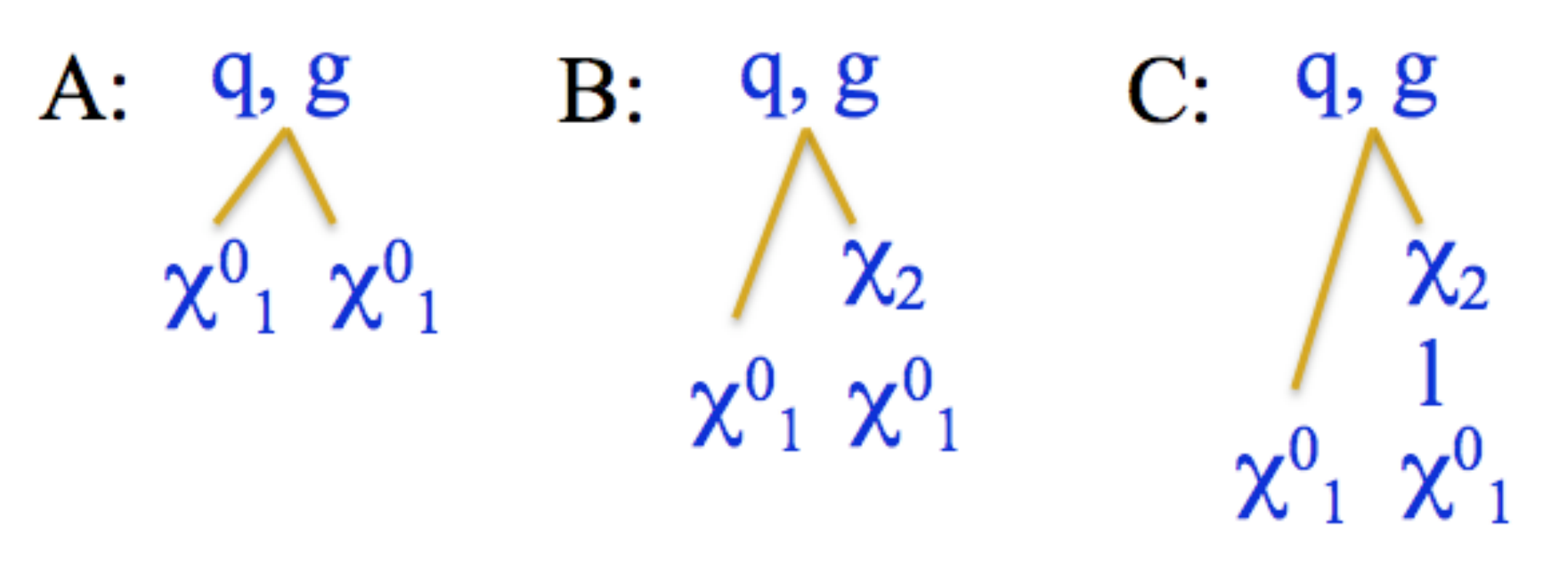}
  \caption{The three dominant combinations of elementary mass spectra forming three classes of eigenmodes denoted A, B and C. Note that the LSP has to be identical for both decay chains.}
  \label{fig:theeigenmodes}
\end{figure*}
Together they cover $98\%$ of the SUSY branching fraction. 
Hence, finding supersymmetry can be reduced, in first order, to the investigation of these three combinations of elementary mass spectra.

In order to investigate them, one has to take into account all the different cases shown in Figure \ref{fig:elementarymassspectraMSSM}.
This gives rise to the eigenmodes listed in Table \ref{tab:theeigenmodes}
which also shows the corresponding final states. 
\begin{table*}
  \centering
  \begin{tabular}{| c | c | l | }
  Eigenmode & Eigenmode Class & Possible Signatures  \\
  \hline
    Aq   & A  &  \MET + (2jets, 3jets, 4jets)  \\
    Ab   & A  &  \MET + (2 b-jets, 3 b-jets, 4 b-jets)  \\
    At    & A  &   \MET + (2 top, 3 top, 4 top)  \\
    \hline
    BZ  &  B  &  \MET + jets + (ll, $\nu\nu$, qq, bb ) \\
    BW  & B &  \MET + jets  + (qq, l$\nu$, bt ) \\	
    \hline
     C0l  & C  & \MET + jets   \\
     C1l &  C  & \MET + jets  + 1l  \\
     C2l &  C  & \MET + jets  + 2l  \\
  \end{tabular}
  \caption{Eigenmodes of classes A, B, C. The BW and BZ modes include the radiation and decay of the neutral and charged MSSM Higgses, respectively.
  l denotes any of the three leptons $e$, $\mu$ or $\tau$ and $q$ denotes any quark besides $b$ and $t$ which are listed separately.}
  \label{tab:theeigenmodes}
\end{table*}
Note that while all different cases of eigenmodes have to be considered to be sensitive to all kinds of mass spectra that nature may have chosen,
not all possible decay channels of the radiated standard model bosons necessarily have to be analyzed.
For the eigenmodes of class B for instance,
one can start by considering the leptonic decays of W and Z.
Adding additional sub-channels will increase the sensitivity.
A subsequent separation of the Higgs cases compared to W/Z may be achieved by considering invariant mass distributions and ratios of the different decay modes.

Note that the mass parameters can be chosen in a way that their effects on the efficiency are independent 
over most of the parameter space, as for instance $m(\tilde{q})-m(\tilde{\chi}_2)$ (controlling the jet $p_T$) and $m(\tilde{l})-m(\tilde{\chi}_1^0)$ (controlling the lepton $p_T$).

In addition to the delta-m variables, 
for the full analysis there is another independent parameter, $m_{\rm SUSY} = {\rm min}(m(\tilde{g}),m(\tilde{q}))$,
which can be investigated by shifting the whole spectrum in mass.
It factors out the main tradeoff between missing energy and production cross section.

When investigating the jet kinematics for the eigenmodes of classes B and C ($\epsilon_{\rm jets_{BC}}$)
a complication compared to the situation for class A (as discussed in Sect. \ref{sec:susyjets}) arises from to the fact that the first gauginos of the two chains now have different masses
giving rise to an additional parameter $m(\tilde{\chi}_2)$ which has to be considered\footnote{In the cases with additional jets from hadronic decays of radiated standard model bosons 
one may take the scenario discussed so far as a reference and investigate
where the additional jets are able to extend the efficiency borders.}.
For alternative trigger paths the efficiency may be written as a sum of terms 
which depend dominantly on only a subset of the mass parameters.
For instance, $\epsilon_{\rm qq} = \epsilon_{jet100}({\rm max}(m(\tilde{q})-m(\tilde{\chi}_1),m(\tilde{q})-m(\tilde{\chi}_2))) + \epsilon_{2jet50}(min(m(\tilde{q})-m(\tilde{\chi}_1),m(\tilde{q})-m(\tilde{\chi}_2)))$, corresponding to the trigger requirement: one jet with $p_T>100$ or two jets with $p_T>50$.\footnote{Note that doing the analysis in this way naturally includes the search for 
mono-jets in the regions where only one of the delta-m values is sufficiently large.}
 
For the class B eigenmodes there is then only one more parameter to be varied: $m(\tilde{\chi}_2)-m(\tilde{\chi}_1)$. 
While large values will lead to boosted bosons, 
going to smaller values will make the channels of the heavier bosons disappear first
and finally lead to quasi stable long lived particles.
All these details are difficult to study in a coherent way by looking at convoluted SUSY model points. 

Similarly, for the class C eigenmodes there are two new parameters $m(\tilde{\chi}_2)-m(\tilde{l})$ and $m(\tilde{l})-m(\tilde{\chi}_1)$
which control the momenta of the radiated leptons.
One may expect the efficiency to factorizes, in first order, for different final states.
For instance: $\epsilon_{\rm C2l} = \epsilon_{\rm jets_{BC}} \times \epsilon_{l}(m(\tilde{\chi}_2)-m(\tilde{l})) \times \epsilon_{l}(m(\tilde{l})-m(\tilde{\chi}_1))$.
Second order effects resulting form jet-electron and electron-jet fake rates as well as isolation effects
can be studied in addition, they can be expected to be independent of $m_{SUSY}$ for instance (for the same jet and lepton kinematics). 

\begin{subsection}{Analysis Channels}
The last question that remains to be answered is which analyses should be performed 
to investigate the eigenmodes of Table \ref{tab:theeigenmodes}.
Their final states can be grouped together as shown in Table \ref{tab:analyzeschanels}. 
As one observes, a small number of analyzes suffice to investigate the efficiency for all the eignemodes\footnote{Note that
their separation is not needed for discovery. However, the measurement of the relative branching ratios of the two spectra contributing to the two lepton channel, for instance, could provide one of the few handles to determine the composition of the gauginos.}.
This gives a total of 9 final states which can be analyzed in a model-independent way as described above, 
i.e. events should be generated for the contributing eigenmodes\footnote{Note that the separate specification of decays for the two produced sparticles 
probably needs some adjustment to the generator programs currently available.}
and the efficiencies determined as a function of the masses. 
The sensitivity in the pMSSM e.g. can then be determined at generator level   
separately for each analysis by comparing it with the standard model background.
Thereto, each eigenmode is weighted with its branching fraction at a given parameter point\footnote{If a complex spectrum contains several realizations of the same elementary spectrum the contributions are summed up, weighted by the efficiencies for the different mass configurations.}, resulting in a conservative estimate in sensitivity. 
When several eigenmodes are analyzed the total sensitivity for a given parameter point increases
as bigger branching fractions are being covered\footnote{Note that there is no interference of the different eigenmodes since each single event results from only one specific eigenmode.}.

Once this has been done, 
one may consider special SUSY parameter points at the border of the experimental efficiency space 
which may profit from the additional investigation of other combinations of the elementary mass spectra.
\begin{table*}
  \centering
  \begin{tabular}{| l | c | l | }
  Signature & Contributing Eigenmodes \\
  \hline
    \MET + jets     &  Aq, BZ, BW, C0l   \\
    \MET + b-jets &  Ab, (BZ, BW)  \\
    \MET + top  &   At, (BW)   \\
    \MET + jets + 1l  &  BW, C1l   \\
    \MET + jets + 2l  &  BZ,  C2l   \\
  \end{tabular}
  \caption{Analysis channels and contributing eigenmodes. Considering each of the three cases $l= e, \mu, \tau$ results in a total of 9 analyses.}
  \label{tab:analyzeschanels}
\end{table*}
\end{subsection}

\end{section}

\begin{section}{Conclusions}
The \bua may be helpful for any given analysis,
and provides a coherent and experiment driven approach for searches for new physics at the LHC.

Since the \bua separates the analysis and interpretation steps it provides an interface between 
experiment and theory, which allows both sides to focus on their specific tasks.
Since the \bua allows to calculate ${\rm N}_{\rm exp}$ in many dimensions with high precision 
it would allow to choose the pMSSM
as a standard for the interpretation of LHC results.

A comparison of the main features of the traditional approach and the \bua is given in Table \ref{tab:comparizon}.
\begin{table*}
  \centering
  \begin{tabular}{| p{3.6cm} | p{3.5cm} | p{4cm} |}
  Feature & Traditional Approach & \bua \\
  \hline
    Aim & Analysis of  SUSY model points & Analysis of elementary mass spectra \\
    Dependent on SUSY model assumptions & Yes & No \\
    Determination of mass spectrum & Difficult from convoluted spectrum & Favored by direct analysis of eigenmodes \\  
    Benchmarks points & Theoretically motivated & Motivated by detector efficiency \\
    Analysis focus & Changing efficiency due to varying contribution from different modes for different SUSY model points & Investigation of detector efficiency as function of new particle masses for each mode \\
    Coverage of full observable phase space & No & Yes \\	
    Interpretation of results & Within chosen model, limited by number of SUSY model points that can be generated and analyzed & Within complete MSSM, covering full phase space accessible to detector sensitivity  \\
    Precision in ${\rm N}_{\rm exp}$ & Low & High \\    
  \end{tabular}
  \caption{Comparison of main features of Traditional and \bua.}
  \label{tab:comparizon}
\end{table*}
If this approach is applied for SUSY analysis at the LHC
it would profit from adjusted interfaces, both to the generator programs to allow the generation of events for a specific eigenmode,
as well as the SUSY model generator to facilitate the mapping from observable space to SUSY model spaces.

Although the particles of the MSSM cover most ways that new particles could interact given the known standard model couplings, 
one could imagine that a similar decomposition into eigenmodes could be performed based on more general BSM models, which 
then could be analyzed following the \bua. 
A basis for the description of such models could be provided by On Shell Effective Theory \cite{Marmoset}.

In focusing on a systematic investigation of the reach in detector efficiency in terms of the physical observables of the new particles 
(and thus extending the reach in SUSY parameter space)
the application of the \bua could increase the chances for discovery of supersymmetry at the LHC 
as well as facilitate the interpretation of the results.
\end{section}

\section*{Acknowledgments}
I would like the thank the following people for insightful discussions and comments: \\
Dong Su, Michael Peskin, JoAnne Hewett and Thomas Rizzo (special thanks for providing the pMSSM parameter points), 
Jay Wacker, 
Giacomo Polesello
and Paul de Jong.

\end{document}